\def\year{2019}\relax
\documentclass[letterpaper]{article} 
\usepackage{aaai19}  
\usepackage{times}  
\usepackage{helvet}  
\usepackage{courier}  
\usepackage{url}  
\usepackage{graphicx}  
\usepackage{amsmath}
\usepackage{amsfonts}
\usepackage{color}

\newcommand{\newcite}[1]{\citeauthor{#1}~\shortcite{#1}}

\begin{document}
%
\title{Combining Long Short Term Memory and Convolutional Neural Network for Cross-Sentence $n$-ary Relation Extraction}
\author{Angrosh Mandya, Danushka Bollegala, Frans Coenen and Katie Atkinson\\
  Department of Computer Science, University of Liverpool\\
  Liverpool, L69 3BX \\
  England \\
  {\tt \{angrosh, danushka.bollegala, coenen, katie\}@liverpool.ac.uk} \\}

\maketitle
\begin{abstract}
We propose in this paper a combined model of Long Short Term Memory and Convolutional Neural Networks  (\textsc{lstm\_cnn}) that exploits word embeddings and positional embeddings for cross-sentence $n$-ary relation extraction. The proposed model brings together the properties of both \textsc{lstm}s and \textsc{cnn}s, to simultaneously exploit long-range sequential information and capture most informative features, essential for cross-sentence $n$-ary relation extraction. The  \textsc{lstm\_cnn} model is evaluated on standard dataset on cross-sentence $n$-ary relation extraction, where it significantly outperforms baselines such as \textsc{cnn}s, \textsc{lstm}s and also a combined \textsc{cnn\_lstm} model. The paper also shows that the \textsc{lstm\_cnn} model outperforms the current state-of-the-art methods on cross-sentence $n$-ary relation extraction.
 
 

\end{abstract}

\section{Introduction}\label{section_1_introduction}

Research in the field of relation extraction has largely focused on identifying binary relations that exist between two entities in a single sentence, known as \textit{intra-sentence relation extraction}~\cite{bach2007survey}.
However, relations can exist between more than two entities that appear across consecutive sentences. For example, in the text span comprising the two consecutive sentences in \textsc{Listing 1}, there exists a ternary relation \textsf{response} across three entities: \emph{EGFR}, \emph{L858E}, \emph{gefitnib} appearing across sentences. 
This relation extraction task, focusing on identifying relations between more than two entities -- either appearing in a single sentence or across sentences, is known as \textit{cross-sentence $n$-ary relation extraction}.

{
\small
{\sc Listing 1: Text span of two consecutive sentences}
\begin{enumerate}
\item [] ``\textit{The deletion mutation on exon-19 of \textbf{EGFR} gene was present in 16 patients, while the \textbf{L858E} point mutation on exon-21 was noted in 10. All patients were treated with \textbf{gefitnib} and showed a partial response.}''
\end{enumerate}
}
This paper focuses on the \textit{cross-sentence  $n$-ary relation extraction} task. 
Formally, let $\{e_1,..,e_n\}$ be the set of entities in a text span $S$ containing $t$ number of consecutive sentences. For example, in the text span comprising 2 sentences ($t=2$) in \textsc{Listing 1} above, given cancer patients with mutation $v$ (EGFR) in gene $g$ (L858E), the patients showed a partial response to drug $d$ (gefitnib). 
Thus, a ternary relation \textsf{response}(\emph{EGFR}, \emph{L858E}, \emph{gefitnib}) exists among the three entities spanning across the two sentences in \textsc{Listing 1}. The entities $e_1,..,e_n$ in the text span can either appear in a single sentence ($t=1$) or multiple sentences ($t>1$). Thus, given an instance defined as a combined sequence of $m$ tokens $\mathbf{x} = x_1, x_2,...,x_m$ in $t$ consecutive sentences and a set of entities  $\{e_1,..,e_n\}$, the cross-sentence $n$-ary relation extraction task is to predict an $n$-ary relation (if exists) among the entities in $\mathbf{x}$.


Cross-sentence $n$-ary relation extraction is particularly challenging compared to intra-sentence relation extraction for several reasons.
Lexico-syntactic pattern-based relation extraction methods~\cite{hearst1992automatic,brin1998extracting,agichtein2000snowball},  have shown to be highly effective for intra-sentence relation extraction.
Unfortunately, such pattern-based relation extraction methods cannot be readily applied to cross-sentence $n$-ary relation extraction because it is difficult to match lexico-syntactic patterns across longer text spans such as covering multiple sentences. 
Features extracted from the dependency parse trees for individual sentences~\cite{culotta2004dependency,bunescu2005shortest,fundel2006relex,xu2015classifying,miwa2016end} have found to be extremely useful for intra-sentence relation extraction.
However, it is non-obvious as how to merge dependency parse trees from different sentences to extract path-based features for cross-sentence relation extraction. 
Moreover, difficulties in coreference resolution and discourse analysis, further complicate the problem of detecting relations among entities across sentences \cite{elango2005coreference}.


The principal challenges for cross-sentence $n$-ary relation extraction arise from 
(a) difficulties in handling long-range sequences resulting from combining multiple sentences,
(b) modeling the contexts of words related to different entities present in different sentences,
and (c) the problem of representing a variable-length text span containing an $n$-ary relation using a fixed-length representation. 
To address these issues, we propose a combined model consisting a Long Short-Term Memory unit and a Convolutional Neural Network (\textsc{lstm\_cnn}) that exploits both word embedding and positional embedding features for cross-sentence $n$-ary relation extraction.
The LSTM is used as the first layer to encode the combined set of sentences representing an $n$-ary relation, thereby capturing  the long-range sequential information. The hidden state representations obtained from the \textsc{lstm} is then used with the \textsc{cnn} to further identify the salient features for relation classification. Our main contributions in this paper can be summarised as follows:

\begin{itemize}
\item [a.] Propose an \textsc{lstm\_cnn} model that exploits word embedding and position embedding features for cross-sentence $n$-ary relation extraction. 
We compare the proposed model against multiple baselines such as \textsc{cnn}, \textsc{lstm} and a combined \textsc{cnn\_lstm} model. 
Experimental results show that the proposed model significantly outperforms all baselines.
\item [b.] Evaluate the proposed model against state-of-the-art (SOTA) for cross-sentence $n$-ary relation extraction on two different benchmark datasets. Results show that the proposed model significantly outperforms the current SOTA methods for cross-sentence $n$-ary relation extraction. 
\end{itemize}

\section{Related Work}
\label{sec_2_related_work}



There is a large body of research on intra-sentence relation extraction \cite{bach2007survey}.
However, our main focus in this paper is on cross-sentence relation extraction.
Therefore, we will limit our discussion below to the cross-sentence relation extraction. Research on cross-sentence relation extraction has extensively used features drawn from dependency trees \cite{swampillai2010inter,quirk2016distant,peng2017cross}, tree kernels \cite{moschitti2013long,nagesh2016exploiting}, and graph LSTMs \cite{peng2017cross}. Further, studies on inter-sentence relation extraction have limited their attention on extracting binary relations present across sentences \cite{swampillai2010inter,quirk2016distant,moschitti2013long,nagesh2016exploiting}. Recently, \citeauthor{peng2017cross} \shortcite{peng2017cross} proposed graph-LSTMs not only to consider binary relations, but also for $n$-ary relations across sentences. 
Although graph LSTMs are useful to model $n$-ary relations across sentences, the process of creating directed acyclic graphs covering words in multiple sentences is complex and error-prone.
It is non-obvious as where to connect two parse trees and the parse errors compound during the graph creation step.
Moreover, co-reference resolution and discourse features used by \newcite{peng2017cross} do not always improve performance of cross-sentence relation extraction.

We present a neural network-based approach that does not rely on heavy syntactic features such as dependency trees, co-reference resolution or discourse features for cross-sentence $n$-ary relation extraction. Although, previous studies have explored LSTMs and CNNs separately for cross-sentence $n$-ary relation extraction, we propose in this paper, a combined model of lstm\_cnn network that simply takes as input the combined sequence of sentences containing $n$-ary relations. While, LSTMs generate features that preserve long-range relations among words in the combined sequence of sentences, CNNs can generate different weighted combinations of those features and select the most informative ones via pooling. 
Although recently several studies have explored combining CNNs and RNNs for various NLP tasks such as text classification \cite{lai2015recurrent,lee2016sequential,hsu2017hybrid} and sentiment analysis \cite{wang2016combination}, to the best of our knowledge, we are the first to propose a combined \textsc{lstm\_cnn} model for cross-sentence $n$-ary relation extraction. 


\section{Cross-Sentence $n$-ary Relation Extraction}\label{sec_3_model}

\begin{figure}[t]
    \centering
    \includegraphics[width=0.47\textwidth]{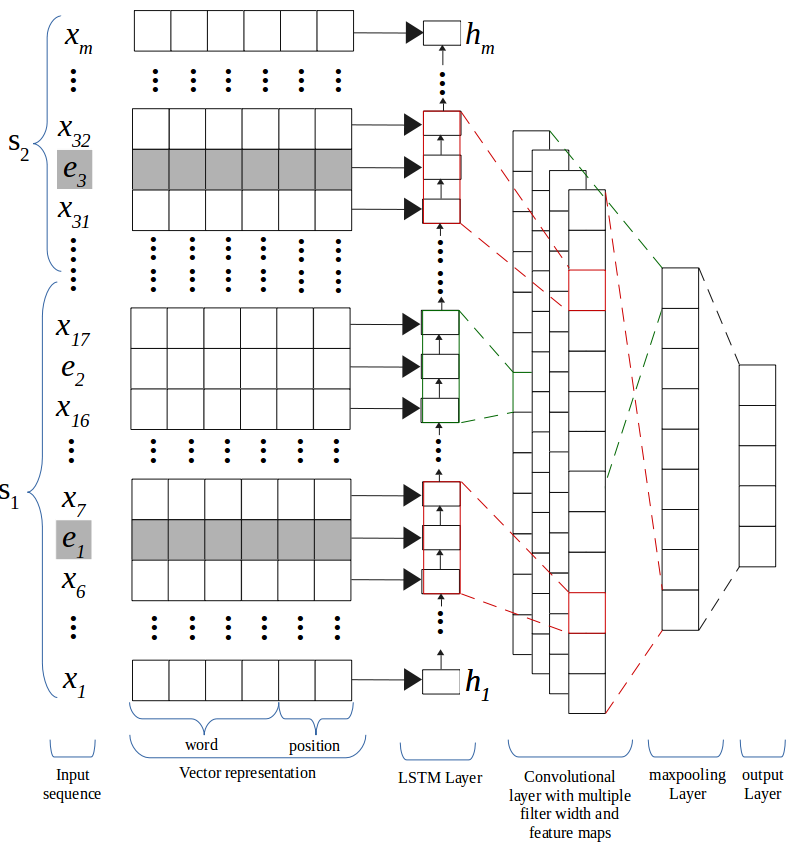}
    \caption{Architecture of the \textsc{lstm\_cnn+wf+pf} model for cross-sentence $n$-ary relation extraction. The input to the network is the sequence of tokens from text span (with two sentences and three entities) shown in \textsc{Listing 1}. The position features are derived for highlighted entities ($e_1$ and $e_3$).}
    \label{fig:lstm_cnn_architecture}
\end{figure}

The architecture of the proposed \textsc{lstm\_cnn+wf+pf} model - combined \textsc{lstm\_cnn} using word features (\textsc{wf}) and positional features (\textsc{pf}) for cross-sentence $n$-ary relation extraction is shown in Figure \ref{fig:lstm_cnn_architecture}. Next, we describe the different components of the proposed model.

\subsection{Input Representation}

The input to the lstm\_cnn model is the combined sequence of tokens in a text span $S$ comprising $t$ consecutive sentences where an $n$-ary relation exists between two entities.
The sequence of tokens is transformed into a combination of word embeddings and position embeddings as follows:

\subsubsection{Word Embeddings}

The transformation of words into lower dimensional vectors are observed to be useful in capturing semantic and syntactic information about words~\cite{mikolov2013efficient,pennington2014glove}. Thus, each of the words in the combined sequence $x = {x_1, x_2,...,x_n}$ is mapped to a $k-$dimensional embedding vector using a look-up matrix $\mathbf{W} \in \mathbb{R}^{|V|\times k}$ where $|V|$ is the number of unique words in the vocabulary.

\subsubsection{Position Features} Following \citeauthor{zeng2014relation} \shortcite{zeng2014relation}, positional features (PFs) are used to encode the position of entities for $n$-ary cross-sentence relation extraction. Given entity mentions $e_1,..,e_n$ in the sequence $x = {x_1, x_2,...,x_n}$, Although $n$ \textsc{pf}s can be defined based on $n$ entities, the proposed model, specifically considers only $e_1$ and $e_n$ to create position embeddings for the input sequence because preliminary experiments show that having $n$ \textsc{pf}s decreases the performance of the model. Thus, the model defines two sets of \textsc{pf}s $PF_1$ and $PF_n$ for the entities $e_1$ and $e_n$, respectively, as a combination of the relative distances from the current word to the respective entity. The position embedding matrices are randomly initialised and the relative distance of words $w.r.t$ entities are transformed into real valued vectors by looking up the position embedding matrices. 

Thus, the vector representation for models using position features, transforms an instance into a matrix $\mathbf{S} \in \mathbb{R}^{s \times d}$ by combining the word embeddings and position embeddings, where $s$ is the sentence length and $d = d^a + d^b \times 2$ ($d^a$ and $d^b$ are the dimensions of respectively the word and position embeddings).

\subsection{LSTM Layer}

Although RNNs are useful in learning from sequential data, these networks are observed to suffer from the problem of exploding or vanishing gradient, which makes it difficult for RNNs to learn long distance correlations in a sequence \cite{hochreiter1997long,hochreiter2001gradient}. To specifically address this issue of learning long-range dependencies, LSTM \cite{hochreiter2001gradient} was proposed which maintains a separate memory cell that updates and exposes the content only when deemed necessary. Given the long-range sequential information resulting from combined set of sentences expressing an $n$-ary relation, LSTM is an excellent choice to learn long-range dependencies. Thus, as shown in Figure \ref{fig:lstm_cnn_architecture}, the transformed vector representation combining word embeddings and position features is provided as input to the LSTM layer. The LSTM units at each time step $t$ is defined as a collection of vectors in $\mathbb{R}^{l}$ and comprises the following components: an input gate $i_t$, a forget gate $f_t$, an output gate $o_t$, a memory cell $c_t$ and a hidden state $h_t$. $l$ is number of LSTM units and the entries of the gating vectors $i_t, f_t$ and $o_t$ are in $[0,1]$. The three adaptive gates $i_t, f_t$ and $o_t$ depend on the previous state $h_{t-1}$ and the current input $x_t$ (Equations  1-3). The candidate update vector $g_t$ (Equation 4) is also computed for the memory cell.

\begin{align}
i_t = \sigma(\mathbf{W}_i x_t + \mathbf{U}_i h_{t-1} + b_i) \\
f_t = \sigma(\mathbf{W}_f x_t + \mathbf{U}_f h_{t-1} + b_f) \\
o_t = \sigma(\mathbf{W}_o x_t + \mathbf{U}_o h_{t-1} + b_o) \\
g_t = \tanh(\mathbf{W}_g x_t + \mathbf{U}_g h_{t-1} + b_g) 
\end{align}

The current memory cell $c_t$ is a combination of the previous cell content $c_{t-1}$ and the candidate content $g_t$, weighted respectively by the input gate $i_t$ and forget gate $f_t$ (Equation 5).

\begin{align}
c_t = i_t \odot g_t + f_t \odot c_{t-1}
\end{align}

The hidden state $h_t$, which is the output of the LSTM units is computed using the following equation:

\begin{align}
h_t = o_t \odot \tanh(c_t).
\end{align}

Here $\sigma$ denotes a sigmoid function and $\odot$ denotes element-wise multiplication.

\subsection{CNN Layer}

Let $h_i \in \mathbb{R}^{l}$ be the $l$-dimensional hidden state vector corresponding to the $i$-th token in the combined sequence $\textbf{x}$.  The combined hidden state vectors in the sequence of length $m$ is represented as:

\begin{equation}
h_{1:m} = h_1 \oplus h_2 \oplus ...\oplus h_m,
\end{equation}

where $\oplus$ denotes vector concatenation. 
In general, let $h_{i:i+j}$ refer to the concatenation of hidden state vectors $h_i, h_{i+1}, ..., h_{i+j}$. The convolution operation involves a filter $\mathbf{w} \in \mathbb{R}^{pl}$, which is is applied to a window of $p$ hidden state vectors to generate a new feature. For instance, a feature $c_i$ is generated from a window of hidden state vectors $h_{i:i+p-1}$.


\begin{equation}
c_i = f(\mathbf{w} \cdot h_{i:i+p-1} + b).
\end{equation}

Here $b \in \mathbb{R}$ is the bias term and $f$ is a non-linear function such as the rectified linear unit (ReLU). This filter is applied to each possible window of hidden state vectors in the combined sequence $h_{1:p},h_{2:p+1},\ldots,h_{m-p+1:n}$ to produce a feature map $c \in \mathbb{R}^{m-p+1}$ given by,
\begin{equation}
c = [c_1, c_2, ..., c_{m-p+1}] .
\end{equation}
Max-pooling is applied over the feature map to take the maximum value $\hat{c}=\max\{c\}$ as the feature corresponding to this particular filter. The use multiple filters and select the most important feature (one with the highest value) for each feature map. Finally, the use of multiple filters with varying window sizes result in obtaining a fixed length vector $\mathbf{g} \in \mathbb{R}^{fw}$, where $f$ is the number of filters and $w$ is the number of different window sizes.

\subsection{Predicting $n$-ary Relations}

The task of predicting $n$-ary relations is modeled both as a binary and multi-class classification problem. The output feature vector $\mathbf{g}$ obtained from the convolution and max-pooling operation is passed to softmax layer, to obtain the probability distribution over relation labels.  Dropout \cite{srivastava2014dropout} is used on the output layer to prevent over-fitting. Thus, given a set of instances, with each instance being a text span $S_i$ comprising $t$ consecutive sentences (combined sequence of tokens $\mathbf{x}=x_1, x_2,...x_m$), entity mentions $e_1,...,e_n$ and having an $n$-ary relation $r$, the 
cross-entropy loss for this prediction is defined as follows:
\begin{equation}
J(\theta) = \sum_{i=1}^s \log p(r_i|S_i, \theta)
\end{equation}
where $s$ indicates the total number of text spans and $\theta$ indicates the parameters of the model.

\subsection{Implementation details}

The proposed model is implemented using Tensorflow \cite{abadi2016tensorflow} and will be made publicly available upon paper acceptance. The hyper-parameters of the models were set based on preliminary experiments on an independent development dataset. Training was performed following mini-batch gradient descent (SGD) with batch size of 10. The models were trained for at most 30 epochs, which was sufficient to converge. The dimensions of the hidden vectors for the LSTM was set to 300. The window sizes for CNN was set to 3,4 and 5, and experiments were conducted with different number of filters set to 10 and 128. 
Word embeddings were initialised using publicly available 300-dimensional Glove word vectors trained on a 6 billion token corpus from Wikipedia and web text \cite{pennington2014glove}. The dimensions for position embeddings was set to 100 and were initialised randomly between [-0.25, 0.25].

\section{Experiments}\label{sec_4_experiments}

\subsection{Datasets}

We conduct experiments using the following datasets.

\subsubsection{Quirk and Poon (\textsc{qp}) Dataset}
We use the dataset\footnote{\url{http://hanover.azurewebsites.net}} developed by \citeauthor{quirk2016distant} \shortcite{quirk2016distant} and \citeauthor{peng2017cross} \shortcite{peng2017cross} for the task of cross-sentence $n$-ary relation extraction. Distant supervision was followed to extract relations involving \textit{drug, gene} and \textit{mutation} triples from the biomedical literature available in PubMed Central\footnote{\url{http://www.ncbi.nlm.nih.gov/pmc}}. The idea of \textit{minimal span} \cite{quirk2016distant} was used to avoid co-occurrence of the same entity triples and also to obtain spans with $\leq$ 3 consecutive sentences to avoid candidates where triples are far apart in the span. A total of 59 drug-gene-mutation triples was used to obtain 3,462 ternary relation instances and 3,192 binary relation instances (involving drug-mutation entities) as positive examples. The dataset has instances with ternary and binary relations, either appearing in a single sentence or across sentences.  Each instances is labeled using four labels: `resistance', `resistance or non-response', `response', and `sensitivity'. The label `none' is used for negative instances. Negative samples were generated by randomly sampling co-occurring entity triples without known interactions, following the same restrictions used for obtaining positive samples. Negative examples were sampled as the same number of positive samples to develop a balanced dataset.

\subsubsection{Chemical Induced Disease (\textsc{cid}) Dataset}
We also evaluate the proposed model using the \textsc{cid} dataset\footnote{\url{https://github.com/JHnlp/BC5CIDTask}}, which provides binary relation instances between chemicals and related diseases. We followed the methodology of \citeauthor{gu2016chemical} \shortcite{gu2016chemical} to obtain relation instances from the corpus. Accordingly, a total of 1206, 1999 and 1330 positive instances were obtained for binary relations in single sentences and total of 702, 788 and 786 positive instances were binary relations across sentences, respectively. Negative instances were created following the same restrictions, however without any known interactions between entities.

\subsubsection{SemEval-2010 Task 8 (\textsc{se}) Dataset}. The SemEval-2010 Task 8 dataset \cite{hendrickx2009semeval} is a standard dataset used intra-sentence relation extraction. The \textsc{se dataset} defines 9 relation types between nominals. The relation `other' is used to denote negative type. The dataset consists of 8,000 training and 2,717 test sentences.

\subsection{Evaluation Metrics}

We conduct five-fold cross-validation and report average test accuracy on held-out folds experiments using \textsc{q\&p dataset}, as prior work \cite{peng2017cross} follow similar evaluation measures. To avoid training and test contamination, held-out evaluation is conducted in each fold, based on categorizing instances related to specific entity pairs (binary relations) or entity triples (ternary relations). For example, for binary relations, the instances relating to the first 70\% of the entity pairs drawn from a unique list of entity pairs are used as training set. Instances relating to the next 10\% and last 20\% are used as development set and test set, respectively. 
For \textsc{cid dataset}, the Precision, Recall and F-score on test set is reported, since the corpus is already divided in train, development and test set and also for comparison as previous studies \cite{gu2016chemical,gu2017chemical,zhou2016exploiting} have used similar measures for reporting the performance. 
For \textsc{se dataset}, we used 10\% of randomly selected instances from the training set as the development set. To evaluate the test set, the official task setting \cite{hendrickx2009semeval} was followed and we report the official macro-averaged F1-Score on the 9 relation types.

\subsection{Baseline models}

The proposed \textsc{lstm\_cnn+wf+pf} model is evaluated against the following baseline models: 
(a) \textbf{\textsc{cnn+wf}}: a \textsc{cnn} model using word features alone; 
(b) \textbf{\textsc{cnn+wf+pf}}: a \textsc{cnn} model using word features and positional features; 
(c) \textbf{\textsc{lstm+wf}}: an \textsc{lstm} model using word features alone; 
(d) \textbf{\textsc{lstm+wf+pf}}: an \textsc{lstm} model using word features and positional features;
(e) \textbf{\textsc{cnn\_lstm+wf}}: a model that begins with a \textsc{cnn} layer followed by an \textsc{lstm} layer and uses word features only;
(f) \textbf{\textsc{cnn\_lstm+wf+pf}}: model that begins with a \textsc{cnn} layer followed by \textsc{lstm} layer and employs word features and position features; (g)  \textbf{\textsc{lstm\_cnn+wf}}: model that begins with an \textsc{lstm} layer followed by a \textsc{cnn} layer and employs word features only.

\subsection{Results and Discussion}



\subsubsection{Performance of the proposed model.} 
The performance of the proposed model \textsc{lstm\_cnn+wf+pf} for cross-sentence $n$-ary relation extraction  on \textsc{q\&p dataset} is shown in Tables \ref{table:drug_var_gene_results} and \ref{table:drug_var_results}. As seen in Tables \ref{table:drug_var_gene_results} and \ref{table:drug_var_results}, the \textsc{lstm\_cnn+wf+pf} model achieves statistically significant accuracy ($p \leq 0.05; \text{Friedman Test}$) against all baseline models such as \textsc{cnn+wf, cnn+wf+pf, lstm+wf, lstm+wf+pf, cnn\_lstm+wf, cnn\_lstm+wf+pf} and \textsc{lstm\_cnn+wf}, for both cross-sentence ternary and binary relation extraction. The results showing the performance of the combined \textsc{lstm\_cnn} model higher than \textsc{cnn} and \textsc{lstm} models in isolation, indicates the usefulness of such combined models for relation extraction. Combining \textsc{lstm} and \textsc{cnn} helps in bringing together the strength of \textsc{lstm}s to learn from long sequences (input sequence) and the ability of \textsc{cnn}s to identify salient features from the hidden-state output sequence from \textsc{lstm} for cross-sentence $n$-ary relation extraction.

Given the above results, it is highly intriguing that a combined model of \textsc{lstm} and \textsc{cnn} using together word features (\textsc{wf}) and positional features (\textsc{pf}), outperforms the evaluated strong baselines. Interestingly, the use of \textsc{wf} alone already helps the combined model (\textsc{lstm\_cnn}) in achieving higher performance against other baselines, particularly for extracting binary relations in single sentences and across sentences, and also ternary relations in single sentences (Tables \ref{table:drug_var_gene_results} and \ref{table:drug_var_results} with $nf=128$). However, it is the addition of \textsc{pf} that helps in drastically improving the performance for relation extraction.  The \textsc{pf} clearly helps the combined \textsc{lstm\_cnn} model by providing useful encoding of the position of words $w.r.t$ entities in the text span, which helps in achieving higher accuracy.

Further, the higher performance achieved in extracting both ternary and binary relations, particularly from cross-sentence text spans which are longer in sequence, indicates that the \textsc{lstm\_cnn+wf+pf} model is highly suitable for extracting relations from longer sequences. Furthermore, the \textsc{lstm\_cnn+wf+pf} model's superior performance extracting ternary and binary relations from single sentences also indicates the suitability of the \textsc{lstm\_cnn+wf+pf} model for relation extraction in single sentences. The evaluation results of the \textsc{lstm\_cnn+wf+pf} on Semeval-2010 Task 8 dataset (standard dataset for intra-sentence relation extraction) presented later in this section, further confirms that the combined model (\textsc{lstm\_cnn}) performs better than employing \textsc{cnn} and \textsc{lstm} in isolation for relation extraction in single sentences.



\begin{table}[t]
\begin{center}
{\small
\begin{tabular}{l c c c c}
\hline
 & \multicolumn{2}{c}{single} & \multicolumn{2}{c}{cross} \\
 & \multicolumn{2}{c}{sentence} & \multicolumn{2}{c}{sentences} \\
\hline
 & $nf$=10 & $nf$=128 & $nf$=10 & $nf$=128  \\ 
\hline

\textsc{cnn+wf} & 72.5 & 75.5 & 75.2 & 76.3 \\
\textsc{cnn+wf+pf} & 73.3 & 73.9 & 78.5 & 78.7 \\
\textsc{lstm+wf}\textsuperscript{$\dagger$} & - & 75.0 & - & 78.2 \\
\textsc{lstm+wf+pf}\textsuperscript{$\dagger$} & - & 74.5 & - &  78.9 \\
\textsc{cnn\_lstm+wf} & 77.6 & 75.4 & 76.9 & 75.3 \\
\textsc{cnn\_lstm+wf+pf} & 72.0 & 53.0 & 76.8 & 62.6 \\
\textsc{lstm\_cnn+wf} & 78.3 & 78.4 & 77.5 & 78.8 \\
\textsc{lstm\_cnn+wf+pf} & 73.1 & \textbf{79.6*} & 80.5 & \textbf{82.9*} \\
\hline

\end{tabular}
}
\end{center}
\caption{\label{table:drug_var_gene_results} Average test accuracy in five-fold cross-validation for \textit{drug-gene-mutation ternary interactions} in \textsc{qp dataset}. $nf$ - number of filters. $\dagger$ \textsc{lstm+wf} and \textsc{lstm+wf+pf} models does not use filters}
\end{table}

\begin{table}[t]
\begin{center}
{\small
\begin{tabular}{l c c c c}
\hline
 & \multicolumn{2}{c}{single} & \multicolumn{2}{c}{cross} \\
 & \multicolumn{2}{c}{sentence} & \multicolumn{2}{c}{sentences} \\
\hline
 & $nf$=10 & $nf$=128 & $nf$=10 & $nf$=128  \\ 
\hline
\textsc{cnn+wf} & 68.9 & 72.4 & 73.2 & 76.6 \\
\textsc{cnn+wf+pf} & 74.0 & 74.2 & 81.3 & 81.3 \\
\textsc{lstm+wf}\textsuperscript{$\dagger$} & - & 75.4 & - & 80.3 \\
\textsc{lstm+wf+pf}\textsuperscript{$\dagger$} & - & 74.4 & - & 80.8  \\
\textsc{cnn\_lstm+wf} & 71.2 & 72.3 & 76.5 & 76.5 \\
\textsc{cnn\_lstm+wf+pf} & 74.7 & 56.2 & 81.2 & 74.4  \\
\textsc{lstm\_cnn+wf} & 74.9 & 76.7 & 79.7 & 82.0 \\
\textsc{lstm\_cnn+wf+pf} & 85.3 & \textbf{85.8*} & 85.1 & \textbf{88.6*}  \\
\hline

\end{tabular}
}
\end{center}
\caption{\label{table:drug_var_results} Average test accuracy in five-fold cross-validation for \textit{drug-gene binary interactions} in \textsc{qp dataset}. $nf$ - number of filters. $\dagger$ \textsc{lstm+wf} and \textsc{lstm+wf+pf} models does not use filters}
\end{table}

\subsubsection{Where exactly does \textsc{lstm\_cnn} model score?} 
To assess the contribution of \textsc{lstm\_cnn+wf+pf} against the baseline models, we divided each dataset into three groups based on the distance between entity $e_1$ and $e_n$ in the text span. Specifically, we calculated the average number of tokens ($\mu$) between $e_1$ and $e_n$ and the standard deviation ($\sigma$) over different lengths of tokens between $e_1$ and $e_n$ in the dataset. Thus, if $k$ is the total number of tokens between $e_1$ and $e_n$, the dataset was divided into the following three groups: 
(a) short-distance spans ($k \! \leq \! \mu \!- \! \sigma$); 
(b) medium-distance spans ($\mu \! - \! \sigma \! < \! k \! < \! \mu \! + \! \sigma $); 
(c) long-distance spans ($k \! \geq \! \mu \! + \! \sigma $). 
Analysing the performance of models on different groups of spans divided in the above manner will provide insights into the model's performance on different sequence lengths and the contribution of different features for relation extraction.

The performance of various models on three groups of sentences, divided based on the number of tokens between entities $e_1$ and $e_n$ in the text span is provided in Table \ref{table:length_of_tokens}. As seen in Table \ref{table:length_of_tokens}, the proposed \textsc{lstm\_cnn+wf+pf} model score higher particularly for medium-distance spans ($\mu \! - \! \sigma \! < \! k \! < \! \mu \! + \! \sigma $) and long-distance spans ($k \! \geq \! \mu \! + \! \sigma $). For example, for short-distance and long-distance spans involving ternary relations across sentences, the \textsc{lstm\_cnn+wf+pf} model predicts ternary relations correctly for a higher percent of $81.3$ and $82.9$ spans, respectively. Similarly, the percentage of correct predictions for binary relation extraction in single sentences and across sentences is significantly higher than the performance of other models. These results clearly indicate that the combined \textsc{lstm\_cnn} model is more useful compared to using \textsc{cnn} and \textsc{lstm} models in isolation for cross-sentence $n$-ary relation extraction, particularly where the distance between the first ($e_1$) and the last entity ($e_2$) is large. In other words the combined \textsc{lstm\_cnn} models are more useful in extracting relations from larger spans of consecutive sentences.

Further, the highest margin between \textsc{lstm\_cnn+wf+pf} and the baselines is recorded for binary interactions in single sentences and across sentences with an accuracy of 85.8 and 88.6, respectively (Table \ref{table:drug_var_results}). This is followed by ternary interactions in single sentences and across sentences with an accuracy of 79.6 and 82.9, respectively (Table \ref{table:drug_var_gene_results}). It is interesting to note that the average length of tokens ($\mu$) between entities in text spans in the datasets relating to binary and ternary interactions in single sentences and across sentences is of the order 19, 29, 34 and 44, respectively. Based on these results, it can be broadly concluded that the contribution of \textsc{pf} decreases with the increase in the distance between entities in the text span.

\begin{table}[t]
\begin{center}
{\small
\begin{tabular}{l | c | c | c }
\hline
Model & \scriptsize{ $k \! \leq \! \mu \!- \! \sigma$} &  \scriptsize{$\mu \! - \! \sigma \! < \! k \! < \! \mu \! + \! \sigma $ } & \scriptsize{$k \! \geq \! \mu \! + \! \sigma $} \\
 		& (\%) & (\%) & (\%) \\
\hline

\multicolumn{4}{c}{\textit{drug-gene-mutation - ternary relations - cross sentence ($\mu$=44)}} \\
\hline
\textsc{cnn+wf} & 82.9 & 74.9 & 79.8 \\
\textsc{cnn+wf+pf} & 84.7 & 76.5 & 80.3 \\
\textsc{lstm+wf} & 46.2 & 77.0 & 79.5 \\
\textsc{lstm+wf+pf} & 54.2 & 77.6 & 80.4 \\
\textsc{cnn\_lstm+wf} & 51.4 & 74.9 & 79.0 \\
\textsc{cnn\_lstm+wf+pf} & 86.2 & 74.8 & 78.8 \\
\textsc{lstm\_cnn+wf} & 52.0 & 76.0 & 79.1 \\
\textsc{lstm\_cnn+wf+pf} & 81.3 & 81.3 & 82.9 \\
\hline

\multicolumn{4}{c}{\textit{drug-gene-mutation - ternary relations - single sentence ($\mu$=34)}} \\
\hline
\textsc{cnn+wf} & 20.0 & 73.1 & 86.6 \\
\textsc{cnn+wf+pf} & 10.0 & 72.0 & 83.4 \\
\textsc{lstm+wf} & 20.0 & 73.5 & 85.8 \\
\textsc{lstm+wf+pf} & 20.0 & 73.0 & 85.6 \\
\textsc{cnn\_lstm+wf} & 20.0 & 76.2 & 87.3 \\
\textsc{cnn\_lstm+wf+pf} & 20.0 & 69.7 & 88.8 \\
\textsc{lstm\_cnn+wf} & 20.0 & 76.8 & 88.0 \\
\textsc{lstm\_cnn+wf+pf} & 20.0 & 79.5 & 86.6 \\
\hline

\multicolumn{4}{c}{\textit{drug-mutation - binary relations - cross sentence ($\mu$=29)}} \\
\hline
\textsc{cnn+wf} & 0.0 & 79.6 & 78.1 \\
\textsc{cnn+wf+pf} & 20.0 & 83.9 & 82.7 \\
\textsc{lstm+wf} & 20.0 & 80.7 & 79.9 \\
\textsc{lstm+wf+pf} & 20.0 & 81.2 & 80.5 \\
\textsc{cnn\_lstm+wf} & 20.0 & 78.0 & 81.3 \\
\textsc{cnn\_lstm+wf+pf} & 20.0 & 84.8 & 87.3 \\
\textsc{lstm\_cnn+wf} & 20.0 & 81.6 & 83.2 \\
\textsc{lstm\_cnn+wf+pf} & 20.0 & 90.9 & 90.2 \\
\hline
\multicolumn{4}{c}{\textit{drug-mutation - binary relations - single sentence ($\mu$=19)}} \\
\hline
\textsc{cnn+wf} & 16.1 & 73.5 & 66.6 \\
\textsc{cnn+wf+pf} & 18.4 & 74.8 & 67.3 \\
\textsc{lstm+wf} & 17.6 & 77.7 & 66.5 \\
\textsc{lstm+wf+pf} & 16.9 & 75.7 & 64.9 \\
\textsc{cnn\_lstm+wf} & 15.3 & 72.7 & 62.5 \\
\textsc{cnn\_lstm+wf+pf} & 19.2 & 76.8 & 65.8 \\
\textsc{lstm\_cnn+wf} & 16.1 & 76.4 & 67.6 \\
\textsc{lstm\_cnn+wf+pf} & 17.6 & 84.9 & 86.5 \\
\hline
\end{tabular}
}
\end{center}
\caption{\label{table:length_of_tokens} Performance of models on different groups of sentences.$k$ - length of tokens between entities $e_1$ and $e_n$, $\mu$ average number of tokens between $e_1$ and $e_n$, $\sigma$ standard deviation over the length of tokens.}
\end{table}

\subsubsection{\textsc{lstm\_cnn} vs. \textsc{cnn\_lstm}.} 
The results shown above clearly indicate that it is more useful to start with an \textsc{lstm} layer followed by \textsc{cnn} layer (\textsc{lstm\_cnn} model) than having a \textsc{cnn\_lstm} model for cross-sentence $n$-ary relation extraction. As seen from Tables \ref{table:drug_var_gene_results} and \ref{table:drug_var_results}, the \textsc{lstm\_cnn} models perform significantly higher than \textsc{cnn\_lstm} models both for ternary and binary relations in single sentences and across sentences. A \textsc{lstm\_cnn} model is more useful in that, it initially learns from the sequential information available in the input, which is further exploited by \textsc{cnn} max-pooling layer to identify salient features. However, in the \textsc{cnn\_lstm} model, the use of \textsc{cnn} layer with max-pooling as the fist component though helps in identifying salient features from the input, the output from the \textsc{cnn} layer does not retain the sequential information. The \textsc{cnn} output feature vector without sequential information when fed to \textsc{lstm} layer, results in poor performance. This indicates that an \textsc{lstm\_cnn} model is more useful than \textsc{cnn\_lstm} model for cross-sentence $n$-ary relation extraction. Further, as the results show, the addition of position embeddings in the \textsc{cnn\_lstm} model (\textsc{cnn\_lstm+wf+pf}) results in poor performance in comparison to the use of word embeddings alone (\textsc{cnn\_lstm+wf}). This is particularly true for ternary relation extraction (Table \ref{table:drug_var_gene_results}). Further as seen in Table \ref{table:drug_var_gene_results}, the use of higher number of filters combining word embeddings and position embeddings, dramatically lowers the performance. This indicates that position embeddings along with higher number of filters are not useful for \textsc{cnn\_lstm} models. However, it is also worthwhile to note that as seen from Table \ref{table:length_of_tokens}, the \textsc{cnn\_lstm+wf+pf} model extracts ternary relations in single sentences for the higher number of long-distance spans (88.8\%), indicating that \textsc{cnn\_lstm} models are useful in certain cases.

\subsubsection{\textsc{cnn} and \textsc{lstm} models.} 
The results provided above clearly shows that, when used in isolation, \textsc{lstm}-based models are more useful for cross-sentence $n$-ary relation extraction, compared to \textsc{cnn}-based models. 
Interestingly, the use of \textsc{pf} helps only longer sequences (accuracy of 78.9 (\textsc{lstm+wf+pf}) vs. 78.2 (\textsc{lstm+wf}) and 80.8 \textsc{lstm+wf+pf}) vs. 80.3 (\textsc{lstm+wf+pf}) scored for ternary relations in drug-mutation-gene (Table \ref{table:drug_var_gene_results}) and drug-mutation (Table \ref{table:drug_var_results}), respectively). However, for shorter sequences, the use of \textsc{pf} results in decrease in accuracy  (accuracy of 74.5 (\textsc{lstm+wf+pf}) vs. 75.0 (\textsc{lstm+wf}) and 74.4 \textsc{lstm+wf+pf}) vs. 75.4 (\textsc{lstm+wf+pf}) scored for binary relations in drug-mutation-gene (Table \ref{table:drug_var_gene_results}) and drug-mutation (Table \ref{table:drug_var_results}), respectively).
The contribution of \textsc{wf} in \textsc{cnn} model significantly improves with the use of higher number of filters, so much so that the model performs better than combining \textsc{wf} and \textsc{pf}. This is particularly true for extracting ternary relations in single sentences (Table \ref{table:drug_var_gene_results}).

\subsubsection{$n$-positional embeddings.} 
Given entities $e_1,..e_n$ in the text span, the proposed \textsc{lstm\_cnn+wf+pf} model employed only $e_1$ and $e_n$ to create positional embeddings. However, we could also create $n$-positional embeddings for each of the $n$ entities in the text span. To this end, we evaluated the \textsc{lstm\_cnn+wf+pf} model using $n$-positional embeddings. The use of $n$-positional embeddings resulted in a lower accuracy of 80.5 and 77.9 (compared to 82.5 and 79.6 using position embeddings for $e_1$ and $e_n$) for ternary relation extraction across sentence and single sentences, respectively. This indicates that using positional embeddings for $e_1$ and $e_n$ is more useful for cross-sentence relation extraction.



\subsubsection{Comparison against the state-of-the-art.} 
As seen from Table \ref{table:q_and_p_comparison}, the proposed \textsc{lstm\_cnnw-wf+pf} model outperforms various state-of-the-art methods for cross-sentence $n$-ary relation extraction on \textsc{q\&p dataset}. These models include \textsc{graph lstm} \cite{peng2017cross}, feature-based models \cite{quirk2016distant}, RNN-based networks such as \textsc{bilstm} \cite{miwa2016end} and \textsc{tree-lstm}, and also combining multi-task learning with \textsc{bilstm} and \textsc{graph lstm} \cite{peng2017cross}. The strength of the proposed model comes from the fact that the previous state-of-the-art methods heavily rely on syntactic features such as dependency tress, co-reference and discourse features, which are time-consuming and less accurate particularly in the biomedical domain. However, in comparison to these models, the proposed \textsc{lstm\_cnn+wf+pf} model does not use any such sophisticated features, but uses much simpler features such as \textsc{wf} and \textsc{pf}. The ability to provide significantly higher performance with much simpler features make the proposed \textsc{lstm\_cnn+wf+pf} an attractive choice for cross-sentence $n$-ary relation extraction.

\begin{table}[t]
\begin{center}
{\small
\begin{tabular}{l c c }
\hline
Model & Single & Cross \\
 & Sent. & Sents. \\
\hline
\hline
\multicolumn{3}{c}{\textit{drug-gene-mutation - ternary relations}} \\
\hline
\hline
\textsc{feature-based} & 74.7 & 77.7 \\
\hline
\textsc{bilstm} & 75.3 & 80.1 \\
\textsc{graph lstm-embed} & 76.5 & 80.6 \\
\textsc{graph lstm-full} & 77.9 & 80.7 \\
\hline
\textsc{bilstm+multi-task} & - & 82.4 \\
\textsc{graph lstm+multi-task} & - & 82.0 \\
\hline
\textsc{lstm\_cnn+wf+pf} (proposed model) & \textbf{79.6} & \textbf{82.9} \\
\hline \hline
\multicolumn{3}{c}{\textit{drug-mutation - binary relations}} \\
\hline \hline
\textsc{feature-based} & 73.9 & 75.2 \\
\hline
\textsc{bilstm} & 73.9 & 76.0 \\
\textsc{bilstm-shortest-path} & 70.2 & 71.7 \\
\textsc{tree-lstm} & 75.9 & 75.9 \\
\textsc{graph lstm-embed} & 74.3 & 76.5 \\
\textsc{graph lstm-full} & 75.6 & 76.7  \\
\hline
\textsc{bilstm+multi-task} & - & 78.1 \\
\textsc{graph lstm+multi-task} & - & 78.5 \\
\hline
\textsc{lstm\_cnn+wf+pf} (proposed model) & \textbf{85.8} & \textbf{88.5} \\
\hline

\end{tabular}
}
\end{center}
\caption{\label{table:q_and_p_comparison} Average test accuracy in five-fold cross validation of the proposed model and SOTA methods on $n$-ary cross-sentence relation extraction (\textsc{q\&p dataset})}
\end{table}

The performance of \textsc{lstm\_cnn+wf+pf} model on \textsc{cid dataset} is provided in Table \ref{table:cid_dataset_comparison}. As seen in Table \ref{table:cid_dataset_comparison}, the \textsc{lstm\_cnn+wf+pf} model achieves statistically significant performance for extracting binary relations from text spans with two sentences ($t=2$) against methods based on supervised learning using linguistic features and maximum entropy models. The \textsc{lstm\_cnn+wf+pf} model also performs well in extracting binary relations in single sentences ($t=2$). The combined \textsc{lstm\_cnn+wf+pf} model achieves higher F-score (0.63) against various SOTA methods\footnote{Note that the SOTA methods listed in Table \ref{table:cid_dataset_comparison} does not use any knowledge base or the development set for learning the model.} on \textsc{cid dataset} as shown in Table \ref{table:cid_dataset_comparison}. The combination of \textsc{lstm\_cnn} provides a slight increase than using \textsc{cnn} and \textsc{lstm} separately on \textsc{cid dataset}. The \textsc{cnn}-based models proposed by \citeauthor{nguyen2018convolutional} \shortcite{nguyen2018convolutional} although achieve a higher recall, they tend to achieve a lower precision. The same is the case with \textsc{cnn+me+pp} \cite{gu2017chemical} and \textsc{cnn} \cite{zhou2016exploiting}. On the other hand, \textsc{lstm}s achieve higher precision, but suffer from poor recall (\textsc{lstm, lstm+svmp} \cite{zhou2016exploiting}). In comparison to \textsc{cnn} models and \textsc{lstm} models, the combined \textsc{lstm\_cnn} achieve a higher precision and at the same time do not lose on recall, resulting in achieving a higher F-score on \textsc{cid dataset}.

\begin{table}[t]
\begin{center}
{\small
\begin{tabular}{p{5.0cm}| p{0.4cm} p{0.4cm} p{0.4cm}}
\hline
Model & P & R & F \\
\hline
\multicolumn{4}{c}{Single sentences (text span where $t \! = \! 1$)} \\
\hline
\textsc{linguistic features} \newline\cite{gu2016chemical} & 0.67 & 0.68 & 0.68 \\
\hline
\textsc{cnn} \cite{gu2017chemical}   & 0.59 & 0.55 & 0.57  \\
\hline
\textsc{lstm\_cnn+wf+pf} (proposed model)  & \textbf{0.69} & \textbf{0.70} & \textbf{0.69} \\
\hline

\multicolumn{4}{c}{Across sentences (text span where $t \! = \! 2$)} \\
\hline
\textsc{linguistic features} \newline\cite{gu2016chemical} & 0.51 & 0.29 & 0.37 \\
\hline
\textsc{maximum entropy} \cite{gu2017chemical}   & 0.51 & 0.07 & 0.11  \\
\hline
\textsc{lstm\_cnn+wf+pf} (proposed model)  & \textbf{0.57} & \textbf{0.57} & \textbf{0.57*} \\
\hline

\multicolumn{4}{c}{Across sentences (text span where $t \! \leq \! 2$)} \\
\hline
\textsc{linguistic features + me} \newline\cite{gu2016chemical} & 0.62 & 0.55 & 0.58 \\
\hline
\textsc{cnn+me} \cite{gu2017chemical} & 0.60 & 0.59 & 0.60  \\
\textsc{cnn+me+pp} \cite{gu2017chemical} & 0.55 & 0.68 & 0.61  \\

\hline
\textsc{cnn} \cite{zhou2016exploiting} & 0.41 & 0.55 & 0.47  \\
\textsc{lstm} \cite{zhou2016exploiting} & 0.54 & 0.51 & 0.53  \\
\textsc{lstm+svmp} \cite{zhou2016exploiting} & 0.64 & 0.49 & 0.56  \\
\textsc{lstm+svm+pp} \cite{zhou2016exploiting} & 0.55 & 0.68 & 0.61  \\

\hline
\textsc{svm} \cite{xu2016cd}   & 0.55 & 0.68 & 0.61  \\

\hline
\textsc{cnn} & 0.54 & 0.69 & 0.61  \\
\textsc{cnn+cnnchar}  & 0.57 & 0.68 & 0.62  \\
\textsc{cnn+lstmchar} \newline \cite{nguyen2018convolutional}  & 0.56 & 0.68 & 0.62  \\

\hline
\textsc{lstm\_cnn+wf+pf} (proposed model)  & \textbf{0.63} & \textbf{0.63} & \textbf{0.63} \\
\hline

\end{tabular}
}
\end{center}
\caption{\label{table:cid_dataset_comparison} Comparison of performance of \textsc{lstm\_cnn+wf+pf} with state-of-the-art models on \textsc{cid dataset}. $t$ = number of sentences, P - precision, R - recall, F - F-score.}
\end{table}

\subsubsection{Performance of \textsc{lstm\_cnn} model on \textsc{se dataset}.} 
To examine the performance of the proposed model on standard relation extraction dataset, the \textsc{lstm\_cnn} model was evaluated on \textsc{se dataset} \cite{hendrickx2009semeval}. The \textsc{lstm\_cnn+wf} and \textsc{lstm\_cnn+wf+pf} models achieved F1-scores of 71.6 and 81.5, respectively. These scores are slightly better than employing \textsc{cnn} with \textsc{wf} to obtain an F1-score of 69.7 and combining \textsc{wf} and \textsc{pf} with \textsc{cnn} to achieve an F1-score of 78.9, further suggesting that combining \textsc{lstm} and \textsc{cnn} is useful for relation extraction. 

\section{Conclusion}

To conclude, we presented in this paper a combined \textsc{lstm\_cnn} model that exploits both word embeddings and position embeddings for the task of cross-sentence $n$-ary relation extraction. The experimental results provided in this paper clearly establish that combining \textsc{lstm}s and \textsc{cnn}s offer the ability to harness together the strength of \textsc{lstm}s to learn from longer sequences and the usefulness of \textsc{cnn}s to learn salient features, vital for cross-sentence $n$-ary relation extraction. The comparison with state-of-the-art results further proves the usefulness of combined \textsc{lstm} and \textsc{cnn} model for cross-sentence $n$-ary relation extraction.

\bibliography{emnlp2018}
\bibliographystyle{aaai}

\end{document}